%% file: paper.tex
\documentclass{sig-alternate}

\usepackage[english]{babel}
\usepackage[utf8]{inputenc}
\usepackage[T1]{fontenc}
\usepackage{enumitem}
\usepackage{microtype}
\usepackage{subfigure}
\usepackage{mathtools}
\usepackage[final]{listings} 
\usepackage[skip=2pt]{caption}
\newcommand{\inlineEnum}[1]{%
  \ifcsname c@#1\endcsname%
  \addtocounter{#1}{1}%
\textbf{\arabic{#1})}~%
\else%
\newcounter{#1}%
\setcounter{#1}{1}%
    \textbf{\arabic{#1})}~%
    \fi%
}

\input{packages}
\begin{document}
\title{Leveraging Secure Multiparty Computation\\ in the Internet of Things}

\numberofauthors{1}
\author{
\alignauthor
Marcel von Maltitz and Georg Carle  \\
       \affaddr{Technical University of Munich}\\
       \affaddr{Boltzmannstra{\ss}e 3}\\
       \affaddr{Garching b. München}\\
       \email{\{lastname\}@net.in.tum.de}
}

\maketitle

\begin{abstract}
  Centralized systems in the Internet of Things---be it local middleware or cloud-based services---fail to fundamentally address privacy of the collected data.
We propose an architecture featuring secure multiparty computation at its core in order to realize data processing systems which already incorporate support for privacy protection in the architecture.
\end{abstract}

\section{Introduction}
Smart environments and smart buildings constitute a vital part of the Internet of Things.
In these contexts, sensors are deployed to gather information about the state of the real-world environment.
This information, in turn, represents the data foundation for services that influence the environment state,
provide insights for inhabitants and interact with them.
Examples for these services are public displays, which give statistical information about the building state,
monitoring services for maintenance personnel and anomaly detection systems which detect incidents and failures.

These and many other services have in common that they do not directly work on the raw data gathered by the sensors.
Instead, they use derived aggregated results by computational preprocessing:
Public displays show diagrams of statistical data, monitoring and anomaly detection services work with events and alerts gained by rules, machine learning or other types of computation.

For mediating the data flow between the sensor platforms---the data sources---and the services---the data consumers---typically a middleware is deployed.
Its purpose encompasses collection and storage of raw data, analysis, processing and finally forwarding the obtained results to the data consuming services.
This middleware can either be a local part of the smart environment but can also be provided as cloud service.

This type of architecture and the corresponding handling of data has severe implications for the privacy of the sensor data:
\inlineEnum{fail}The middleware acts as a third party which gains full access to raw data coming from the sensors.
This third party might not even be under control of the administrators of the smart environment and hence untrustworthy.
\inlineEnum{fail}
By pushing data to a third party, sources lose insights into how their data is used afterwards.
Data processing becomes intransparent for them.
\inlineEnum{fail} Similarly, sources lose control over the usage of their data.
Especially, revocation of data requires trust in the data holder to actually obey.
\inlineEnum{fail} Even if trustworthy, the third party is still a high value target for attackers.

\section{Privacy Preserving Data Processing}
Our vision is to realize the described functionality while fundamentally providing privacy protection on the architectural level.
We propose that raw data created by the distributed sources is not collected by a middleware but remains distributed on these sources.
This allows secure computations and can make consent and cooperation of the sources a necessity for the execution.

Our understanding of privacy and data protection is based on \cite{Datenschutzschutzziele,Hansen2015}.
They most importantly feature the protection goals of
\emph{data minimization},
\emph{unlinkability},
\emph{transparency}, and
\emph{intervenability}.
Against this background, the positive implications of our approach are as follows:
The amount of data in the system is minimized since there are no intermediaries which can also access data.
Logically, the derived results are directly transmitted from the sources to the final consumers.
The potential for data misuse and unauthorized recombination of data is decreased since data of different sources is not stored at the same logical place in a linkable fashion.
Specifically, only making allowed computations technically possible concomitantly realizes purpose binding.
The required cooperation of the data sources in turn provides them with information about the ongoing computations and the usage of their data.
This constitutes transparency, especially when this feedback is enhanced with meta information about the final consumer.
Persisting these insights can additionally realize accountability.
Lastly, given the cooperation requirement and aforementioned transparency, they remain in control since they can specifically decide beforehand whether to cooperate and to provide their data for the usage in question or not.

%

\section{Architecture Design}
The provided vision satisfies several data protection goals which are not yet fulfilled by state-of-the-art architectures.
In order to realize this vision technically, the following main challenge has to be addressed:
It must be possible to derive computation results from raw data of different sources without sharing this data among them nor handing it to a third party for computation.

For this purpose, secure multiparty computation (SMC) \cite{Yao1982, Yao1986, SMCBook} can be employed.
Instead of local computations of a third party a secure protocol among the sources is executed \cite{Canetti1999, Canetti2001}.
Afterwards each only knows its own input and the final output of the computation.
All exchanged intermediary data due to technical reasons does not allow recovering other parties inputs.
Mathematical foundations for realizing arbitrary functions as SMC invocations are known since the 80's \cite{BGW88,Chaum1988a} but protocol improvements for security and performance~\cite{BristolMPC,MASCOT} and new applications \cite{Bonawitz2017} are still current research.

For sucessfully applying SMC in smart environment we propose the following architecture:
The formerly stated middleware is replaced by a \emph{gateway}.
The vital difference is that the gateway does not obtain access to the raw data of the sources.
Instead, facing the sources it only fulfills management and orchestration purposes to carry out SMC computations.
Towards the consumers, it presents an API which abstracts from SMC and resembles an interface a centralized middleware would provide.

\paragraph{Robust automated execution of SMC~\cite{vonMaltitz2018b}}
The main purpose of the gateway is to handle SMC sessions in cooperation with the sources.
For this, the gateway must be initially known by them.
Similarly, upon connection interruption or due to churn of mobile sources a present gateway has to be redetected.
This is realized by a service discovery technology like mDNS~\cite{RFC6762,RFC6763}. 
After detection, a setup between the new source and the gateway is performed:
The gateway is informed about data and computation protocols provided by the new source.
This data constitutes a state about currently obtainable insights about the environment in the form of a metadata directory.
Furthermore, the gateway establishes a control channel to the source allowing to prepare and orchestrate SMC sessions.

The gateway specifies all aspects of an upcoming session and communicates them to the participating sources:
The identity and the connection endpoints of cooperators, the data to be used for computation and the protocol to be executed.
The computation itself is monitored by the gateway.
On success, the gateway receives the result.
If the computation fails, the gateway tries to recover or to fully restart the session.
This is hidden from any consumer in any possible cases to achieve service character.

\paragraph{Data Requests and Access Control}
The purpose of the gateway towards the consumers is to mimic a standard middleware providing data upon request.
Here, the metadata directory provides information to the consumers what data is obtainable at this point in time.
This metadata should also abstract from SMC specifics allowing
to post requests which already declare the aggregation result, \eg ``the average amount of individuals in floor 3.A of the building per hour''.
Receiving these requests, the gateway then transforms them into a corresponding SMC session and replies with the result afterwards.

Correct representation of data requests supports access control, transparency and intervenability essentially.
We assume requests to be authenticated and integrity protected.
The gateway is then able to perform access control and plausbility checks when examining the purpose of the request, the identity of the consumer and the type of requested data.
During SMC session setup the gateway also transmits the original request of the consumer to each collaborator, consequently realizing request transparency for sources.
Additional persisting the requests provides distributed request accountability.
Lastly, this information can be evaluated by the sources before executing the computation.
Each source can decide individually whether to contribute to the requested computation or not.
In case a single source veotes against the computation,
it cannot be executed; this is handled as a special, expected error by the gateway and can be addressed accordingly by it.
In summary, we deliberately leverage the necessity of cooperation when performing computations to support the mentioned further privacy properties.

\section{Conclusion}
We presented a vision of privacy preserving data processing in dynamic environments.
Our design features a management and orchestration middleware for secure multiparty computation
which allows application of SMC as an adaptive and robust service.
Furthermore, we show how the features of SMC can be complemented in order to fulfill further established privacy protection goals.

We see that fundamental innovation in system architecture allows more straightforward addressing of privacy goals.
While also raising new challenges to be solved, they provide an alternative approach to establishing privacy
as an afterthought in a predetermined system.

\clearpage
\section{Acknowledgements}
This work has been supported by the German Federal Ministry of Education
and  Research,  project  DecADe,  grant  16KIS0538  and  the  German-French Academy for the Industry of the Future.
\bibliographystyle{IEEEtran}
\bibliography{literature}

\end{document}

%% file: packages.tex
\usepackage[english]{babel}

\usepackage{url}

\newcommand{\hairspace}{\hspace{1pt}}
\newcommand{\eg}{\mbox{e.\hairspace{}g}.\ }  

\usepackage{amssymb}

\usepackage{tikz}
\usetikzlibrary{shapes}
\usetikzlibrary{calc}
\tikzset{myptr/.style={-{Latex[scale=1.5]}}}

\tikzset{MyRoundedBox/.style={
		draw,
		rounded corners=3pt,
		inner sep=5pt,
		align=center
	}
}

\usepackage{listings}
\usepackage{cite}

\usepackage[binary-units=true]{siunitx}

\mathchardef\mhyphen="2D 

\usepackage{booktabs}
\usepackage{array}
\usepackage{multirow}

\usepackage{pifont}
\usepackage{expdlist}

\usepackage{fancyvrb}